\documentclass[aps,pra,twocolumn]{revtex4}
\usepackage[tbtags]{amsmath}
\usepackage{amssymb,amsfonts,bm,mathptmx,enumerate}
\usepackage{graphicx}

\renewcommand{\vec}[1]{\mathbf{#1}}

\DeclareSymbolFont{symbols}{OMS}{cmsy}{m}{n}

\begin{document}

\title{Casimir effect for tachyonic fields}

\author{Marcin Ostrowski}
\email{m.ostrowski@merlin.fic.uni.lodz.pl}

\affiliation{Department of Theoretical Physics, University of
  {\L}{\'o}d{\'z}, ul.\ Pomorska 149/153, 90-236 {\L}{\'o}d{\'z}, Poland}

\begin{abstract}
In this paper we examine Casimir effect in the case of tachyonic field,
which is connected with particles with negative four-momentum square
($m^2<0$). We consider here only the case of the one dimensional, scalar
field. In order to describe tachyonic field, we use the absolute
synchronization scheme preserving Lorentz invariance. The renormalized
vacuum energy is calculated by means of the Abel-Plana formula. Finally,
the Casimir energy and Casimir force as the functions of distance are
obtained. In order to compare the resulting formula with the standard one,
we calculate the Casimir energy and Casimir force for massive, scalar
field ($m^2>0$).
\end{abstract}

\maketitle

\section{Introduction}

Tachyons are hypothetical particles which move faster than light. Recently,
a number of papers was published, suggesting that tachyons can play a role
in cosmology, as well as in neutrino physics \cite{Pierwszy, Drugi}. In
order to describe the kinematics of this kind of particles, formalism based
on the absolute synchronization scheme should be used. Now, we briefly
describe main results related to the absolute synchronization scheme.
Derivation of these results and exhaustive description can be found
in \cite{Trzeci, Czwarty}.

The main idea is based on the well known fact, that the definition of the
time coordinate depends on the procedure used to synchronize clocks.
Furthermore, the form of Lorentz transformations depends on the synchronization
scheme. In absolute synchronization a new concept, the preferred frame
arise. It is intimately connected with tachyons \cite{Piaty}. Possibly, this
distinguished inertial frame can be identified with cosmic background
radiation frame.

In the formalism presented here, the transformation of coordinates between
inertial frames takes the following form:
\begin{equation}
\label{eq:lt1}
x'(u')=D(\Lambda,u)x(u),
\end{equation}
where $\Lambda$ is an element of Lorentz group, $u$ is the four-velocity of
the preferred frame, and $D(\Lambda,u)$ is a matrix depending on $\Lambda$
and $u$.

Transformation law for the four-velocity of the preferred frame, which
follows from Eq.~\eqref{eq:lt1}, is defined by
\begin{equation}
\label{eq:lt2}
u'=D(\Lambda,u)u.
\end{equation}
Matrix $D(\Lambda,u)$ for any rotation $R$ is given by
\begin{equation}
\label{eq:lt3}
D(R,u)=
\begin{pmatrix}
1&0\\0&R
\end{pmatrix},
\end{equation}
while for the boosts $w$ it takes the form
\begin{equation}
\label{eq:lt4}
D(w,u)=
\begin{pmatrix}
{\tfrac{1}{w^0}}&0\\{-\vec{w}}&
{I+\tfrac{\vec{w}\otimes\vec{w^T}}{1+\sqrt{1+\vec{w}^2}}
-u^0\vec{w}\otimes\vec{u^T}}
\end{pmatrix},
\end{equation}
where $w$ is the four-velocity of the system $x'$ with respect to the system
$x$. Hence, in the absolute synchronization scheme, we get the Lorentz
transformation law for time coordinate in the form
\begin{equation}
x'^{0}=x^0/W^0.
\end{equation}
So, in the absolute synchronization scheme, time coordinate $x^0$ is only
rescaled by some positive factor $1/W^0$. It is the most important property, which
leads to preservation of the absolute causality.

The transformation law for velocities in the absolute synchronization
scheme, derived from Eqs.~\eqref{eq:lt1}, \eqref{eq:lt2} and \eqref{eq:lt4}
is given by
\begin{equation}
\label{eq:pv1}
\vec{v}'=W^0\biggl[\vec{v}+\vec{W}\biggl(\tfrac{\vec{Wv}}
{1+\sqrt{1+\vec{W}^2}}-u^0\vec{uv}-1\biggr)\biggr].
\end{equation}
Contrary to the standard procedure of synchronization, the transformation
law (\ref{eq:pv1}) is well defined not only for subluminal velocities but
also for superluminal. It is continous, and does not produce
`transcendental' tachyons moving with $\vert\vec{v}\vert=\infty$. This
property and preservation of absolute causality make absolute
synchronization scheme useful for description of tachyons.

The line element $ds^2=g_{\mu\nu}(u)dx^{\mu}dx^{\nu}$ is invariant under
Lorentz transformations if the covariant metric tensor is expressed as
follows:
\begin{equation}
g_{\alpha\beta}(u)=\begin{pmatrix}1&u^0\vec{u}^T\\
u^0\vec{u}&-I+\vec{u}\otimes\vec{u}^T(u^0)^2
\end{pmatrix},\end{equation}
where $u=(u^0,\vec{u})$ is four-velocity of the preferred frame.
Contravariant metric tensor is given by
\begin{equation}
g^{\alpha\beta}(u)=\begin{pmatrix}(u^0)^2&u^0\vec{u}^T\\
u^0\vec{u}&-I
\end{pmatrix}.\end{equation}
In the case of tachyons, square of the four-momentum
\begin{equation}
k_{\alpha}k^\alpha=-m^2,
\end{equation}
leads to the dependence of the energy $k_0$ on momentum
$(k_1,k_2,k_3)=\vec{\underline{k}}$ in the form
\begin{equation}
\label{eq:ekov}
k_0=(u^0)^{-1}\biggl[-\vec{u\underline{k}}+
\sqrt{(\vec{u\underline{k}})^2+\vec{k}^2-m^2}\biggr],
\end{equation}
whereas contravariant energy $k^0$ is
\begin{equation}
\label{eq:ekon}
k^0=u^0\sqrt{(\vec{u\underline{k}})^2+\vec{\underline{k}}^2-m^2}.
\end{equation}
Sign of $k^0$ is positive and Lorentz invariant not only for massive
particles but also for tachyons. Values of momentum $\vec{k}$ fulfil
the condition in the form
\begin{equation}
\label{eq:war}
\vec{\underline{k}}^2>\frac{m^2}{1+(\vec{u\underline{k}/
\vert\underline{k}\vert})^2}.
\end{equation}
Values of $\vec{\underline{k}}$ lie outside of the oblate spheroid with
half-axes $m$, $mu^0$ and with the symmetry axis parallel to $\vec{u}$.

A more exhaustive discussion of the absolute synchronisation scheme in the
language of frame bundles is given in \cite{Czwarty}. Another appllications
in physics can be found in \cite{PiecA} (EPR correlations) and \cite{PiecB}
(thermodynamics).

\section{One dimensional tachyonic field}

The quantum description of free tachyonic field was given in \cite{Trzeci}.

In this article we consider only scalar field in one space-dimension.
In this case Eqs.~\eqref{eq:ekov} and \eqref{eq:ekon} take the form
\begin{align}
\label{eq:ekov1}
k_0=&(u^0)^{-1}\biggl[-u^1k_1+\tfrac{1}{u^0}\sqrt{k_1^2-(u^0m)^2}\biggr],\\
k^0=&\sqrt{k_1^2-(u^0m)^2}.
\end{align}
Whereas the condition in Eq.~\eqref{eq:war} reduces to
\begin{equation*}
k_1^2>\frac{m^2}{1+(u^1)^2},
\end{equation*}
which, with the help of the identity $(1+{u^1}^2)(u^0)^2=1$, leads to the
condition
\begin{equation}
\label{eq:pcon}
k_1^2>(mu^0)^2.
\end{equation}
The Lagrangian density of the scalar field can be written in the following
form:
\begin{multline}
\mathcal{L}=\tfrac{1}{2}(g^{\alpha\beta}\partial_\alpha\phi\partial_\beta\phi
+m^2\phi^2)=\\=\tfrac{1}{2}\biggl[(u^0)^2(\partial_0\phi)^2+
u^0u^1(\partial_0\phi\partial_1\phi+\partial_1\phi\partial_0\phi)+\\
-(\partial_1\phi)^2+m^2\phi^2\biggr],
\end{multline}
and canonical momentum is
\begin{equation}
\pi=\frac{\partial\mathcal{L}}{\partial(\partial_0\phi)}=
(u^0)^2\partial_0\phi+u^0u^1\partial_1\phi.
\end{equation}
Hamiltonian density is given by
\begin{equation}
\mathcal{H}=\frac{1}{2}\biggl[(u^0)^2(\partial_0\phi)^2+
(\partial_1\phi)^2-m^2\phi^2\biggr].
\end{equation}
Scalar field in one dimension is represented by the operator:
\begin{equation}
\label{eq:fdek}
\phi(x,u)=\tfrac{1}{\sqrt{2\pi}}\int d\mu(k)(e^{ikx}a^+(k)+e^{-ikx}a(k)),
\end{equation}
where $d\mu(k)$ is an invariant measure defined by
\begin{equation}
\label{eq:miar}
d\mu(k)=dk_0dk_1\Theta(k^0)\delta(k^2+m^2)=
dk^0dk_1\frac{\delta(k^0-\omega_k)}{2\omega_k},
\end{equation}
while the coefficient $\omega_k$ is given by
\begin{equation}
\omega_k=\sqrt{k_1^2-(u^0m)^2}.
\end{equation}
By means of (\ref{eq:miar}), the Fourier decomposition of the field 
$\phi(x,u)$ takes the form
\begin{equation}
\phi(x,u)=\frac{1}{\sqrt{2\pi}}\int_\Gamma\frac{dk_1}{2\omega_k}
\bigl(e^{ikx}a^+(k_1,u)+e^{-ikx}a(k_1,u)\bigr),
\end{equation}
where the integration range $\Gamma$ is determined by Eq.~\eqref{eq:pcon}.

\section{Casimir effect for tachyonic field in the preferred frame}

Now, let us consider the one dimensional, scalar, tachyonic field in the limited
region of length $a$ (Fig.~1) with Dirichlet boundary conditions:
\begin{equation}
\label{eq:bkon}
\phi(x^0,0)=\phi(x^0,a)=0.
\end{equation}

In the frame moving relative to preferred frame, energy of the particle
Eq.~\eqref{eq:ekov1} is not an even function of momentum $k_1$. Hence,
two waves which form the standing wave (and moving in different directions)
have different length and speed. This fact causes some difficulties in
calculations. In order to avoid these difficulties, in this section only
Casimir energy in preferred frame ($u^0=1$) is computed. 

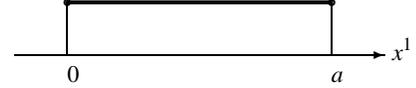
\begin{figure}[h]
\begin{picture}(200,40)
\thicklines
\put(50,35){\circle{2}}
\put(50,35){\line(1,0){100}\circle{2}}
\thinlines
\put(30,15){\vector(1,0){140}}
\put(50,15){\line(0,1){20}}
\put(150,15){\line(0,1){20}}
\put(173,13){\text{$x^1$}}
\put(50,5){\text{0}}\put(150,5){\text{$a$}}
\end{picture}
\caption{One dimensional field.}
\end{figure}

Taking into account the boundary condition Eq.~\eqref{eq:bkon} we obtain
the Fourier expression of the operator $\phi$ such that
\begin{align}
\label{eq:fiop}
\phi(x^0,x^1)=\frac{i}{\sqrt{a}}\sum_{k_n^2>(mu^0)^2}
\frac{\sin{k_nx^1}}{2\omega_k}\bigl[e^{-ik_0x^0}a_n^+-e^{ik_0x^0}a_n\bigr],
\end{align}
where $k_n={\pi}n/a$. The summation in Eq.~\eqref{eq:fiop} is over every $n$
which fulfill condition \eqref{eq:pcon}. Operators $a_n$ and $a_n^+$ are
annihilation and creation operators which satisfy well known commutation
relations:
\begin{align}
[a_n,a_{n'}]&=[a_n^+,a_{n'}^+]=0,\\
[a_n,a_{n'}^+]&=2\delta_{n,n'}\omega_n.
\end{align}
Hamiltonian of tachyonic field (in preferred frame) can be expressed as
\begin{equation}
\label{eq:ham}
H=\frac{1}{2}\int_0^adx\bigl[(\partial_0\phi)^2+(\partial_1\phi)^2-
m^2\phi^2\bigr].
\end{equation}
On inserting operator $\phi$ from Eq.~\eqref{eq:fiop} with its derivatives
into Eq.~\eqref{eq:ham} and integrating with respect to variable $x^1$, we
find
\begin{multline}
H=\sum_{k_n^2>m^2}\tfrac{1}{16}\omega_k^{-2}
(k_0^2-m^2+k_n^2)(a_n^+a_n+a_na_n^+)=\\
=\sum_{k_n^2>m^2}\tfrac{1}{8}(a_n^+a_n+a_na_n^+),
\end{multline}
If we calculate vacuum energy as an expectation value of $H$, we get
\begin{equation}
\label{eq:vacE}
E=\langle{0}{\vert}H{\vert}0\rangle=
\tfrac{1}{2}\sum_{k_n^2>m^2}\sqrt{k_n^2-m^2},
\end{equation}
where $\vert{0}\rangle$ is a vacuum state, defined by equation
$a_n\vert{0}\rangle=0$. It is well defined and stable state (See 
\cite{Trzeci}).

If we compare Eq.~\eqref{eq:vacE} with Eq.~\eqref{eq:ekov1} we can see that the
vacuum energy is equal to the sum of energies of each mode which satisfies
boundary conditions and condition (\ref{eq:pcon}).

After changing the limits of summation, Eq.~\eqref{eq:vacE} can be written as
\begin{equation}
\label{eq:VacEn}
E(a)=\sum_{n=n_0}^{\infty}\sqrt{\biggl(\frac{\pi{n}}{a}\biggr)^2-m^2},
\end{equation}
where $n_0=[ma/\pi]$. (Function $[x]$ gives the smallest integer
greater than or equal to $x$.) The range of summation depends on the
parameter $a$. When it increases, the more low frequencies are rejected by
function $[$ $]$. 

When $a$ goes to infinity we have
\begin{equation}\label{eq:VacEm}
\begin{split}
\sum_{n=n_0}^{\infty}\sqrt{\biggl(\frac{\pi{n}}{a}\biggr)^2-m^2}&
\xrightarrow{a\rightarrow\infty}\frac{a}{\pi}\int_{\tfrac{n_0\pi}{a}}^\infty
dk\sqrt{k^2-m^2}=\\=\frac{a}{\pi}\int_m^\infty&{dk}\sqrt{k^2-m^2}.
\end{split}
\end{equation}
The last stage follows as a result of obvious equation
$\lim_{x\rightarrow\infty}[x]/x=1$.

The vacuum energy (Eq.~\ref{eq:VacEn}), also in the case of
tachyonic fields, is infinite. Therefore we introduce the renormalized
`Casimir energy' as a difference between vacuum energy in the two
configurations shown in Fig.~2. Parameter $L$ is the total length of
considered region.
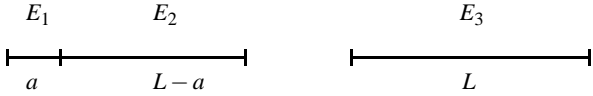
\begin{figure}[h]
\begin{picture}(250,40)
\thicklines
\put(10,15){\line(1,0){90}}
\put(140,15){\line(1,0){90}}
\put(10,12){\line(0,1){6}}
\put(30,12){\line(0,1){6}}
\put(100,12){\line(0,1){6}}
\put(140,12){\line(0,1){6}}
\put(230,12){\line(0,1){6}}
\put(17,3){\text{$a$}}\put(65,3){\text{$L-a$}}
\put(182,3){\text{$L$}}
\put(17,29){\text{$E_1$}}\put(65,29){\text{$E_2$}}
\put(181,29){\text{$E_3$}}
\end{picture}
\caption{Two configurations of boundary conditions.}
\end{figure}
In this case the Casimir energy is defined by
\begin{equation}
E_c(a)=\lim_{L\rightarrow\infty}(E_1(a)+E_2(L-a)-E_3(L)).
\end{equation}
In order to calculate energy $E_1$ we use Eq.~\eqref{eq:VacEn}, while
energies $E_2$, $E_3$ should be calculated with the help of
Eq.~\eqref{eq:VacEm}. Hence,
\begin{multline}
E_c(a)=\\
=\biggl(\sum_{n=n_0}^{\infty}\sqrt{\bigl(\pi{n}/a\bigr)^2-
m^2}-\frac{a}{\pi}\int_m^\infty{dk}\sqrt{k^2-m^2}\biggr).
\end{multline}
Introducing parameter $\lambda=am/\pi$ and introducing the new
variable of integration $n=ak/\pi$ we get
\begin{equation}
\label{eq:CasE}
E_c(a)=\frac{\pi}{a}\biggl(\sum_{n=n_0}^\infty\sqrt{n^2-
\lambda^2}-\int_\lambda^\infty{dn}\sqrt{n^2-\lambda^2}\biggr),
\end{equation}
where $n_0=[am/\pi]=[\lambda]$.

To compute the expression mentioned above we use the Abel-Plana formula
\begin{equation}
\label{eq:ABF}
\sum_{n=0}^{\infty}f(n)-\int_{0}^{\infty}f(n)dn=\tfrac{1}{2}f(0)
+i\int_{0}^{\infty}dt\frac{f(it)-f(-it)}{e^{2\pi{t}}-1}.
\end{equation}
This formula is often applied in renormalization problems. Its applications
and generalizations can be found in many works. For example, we mention
here \cite{Szosty, Osmy}.

We can write Eq.~\eqref{eq:CasE} in following form:
\begin{multline}
E_c(a)=\frac{\pi}{a^2}\biggl[\sum_{n=0}^\infty\sqrt{(n+n_0)^2-
\lambda^2}+\\
-\int_0^\infty\sqrt{(n+n_0)^2-\lambda^2}dn-\int_\lambda^{n_0}
\sqrt{n^2-\lambda^2}dn\biggr].
\end{multline}
Furthermore, applying of the formula (\ref{eq:ABF}) we finaly get
\begin{multline}
\label{eq:CasE2}
E_c(a)=-\frac{\pi}{a}\\
\times\biggl[\sqrt{2}\int_0^{\infty}dt
\frac{\sqrt{\sqrt{(n_0^2-\lambda^2-t^2)^2+4n_0^2t^2}-n_0^2+\lambda^2+t^2}}
{e^{2\pi{t}}-1}+\\+
\tfrac{1}{2}\lambda^2\ln\biggl(\frac{n_0+\sqrt{n_0^2-\lambda^2}}{\lambda}
\biggr)+\tfrac{1}{2}(n_0-1)\sqrt{n_0^2-\lambda^2}\biggr].
\end{multline}
The numerical computation of the expression (\ref{eq:CasE}) are shown in
Figs~3-6.

\begin{figure}[h]
\begin{center}
\includegraphics[width=7cm]{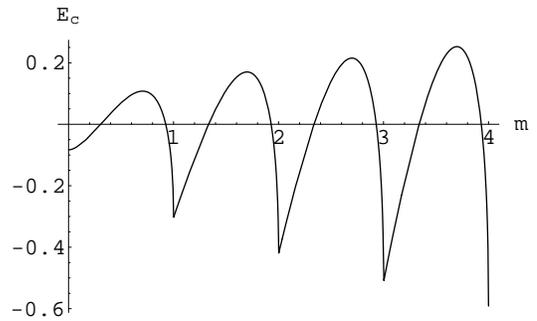}
\end{center}
\caption{Casimir energy of tachyonic field as a function of mass $m$,
 corresponding to $a=\pi$.}
\end{figure}

The Fig.~3 shows the Casimir energy (Eq.~\ref{eq:CasE}) as a function of
mass $m$. Presented curve corresponds to distance $a=\pi$ and $u^0=1$.
We obtain here the function piecewise differentiable, the points of
indifferentiability are for integer values of parameter $\lambda$. This
discontinuity of derivative is related to the existence in expressions
discontinuous parameter $n_0=[\lambda]$.

\begin{figure}[h]
\begin{center}
\includegraphics[width=7cm]{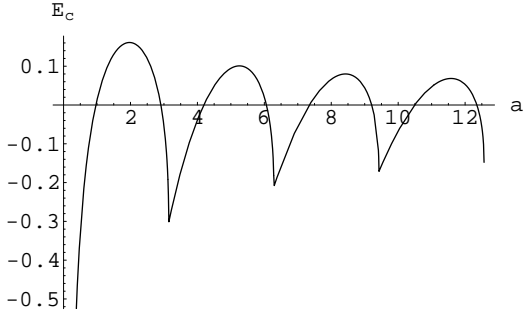}
\caption{Casimir energy of tachyonic field as a function of the distance $a$,
corresponding to $m=1$.}
\end{center}
\end{figure}

The Fig.~4 shows Casimir energy (Eq.~\ref{eq:CasE}) as a function of the
parameter $a$ (Recall that $a$ is the distance between two points where
$\phi$ vanishes). The presented curve corresponds to $m=1$ and $u^0=1$
(preferred frame).

Since the dependence on $m$ enters to the expression \eqref{eq:CasE} only
through parameter $\lambda$, therefore for other values of $m$ the Casimr
energy function will have the same character, only its horizontal and
vertical scale will change.

\begin{figure}[h]
\begin{center}
\includegraphics[width=7cm]{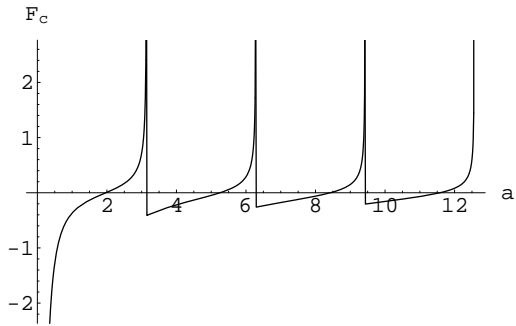}
\caption{Casimir force for tachyonic field as a function of the distance
 $a$, corresponding to $m=1$.}
\end{center}
\end{figure}

The Fig.~5 shows Casimir force as a function of the distance $a$, in
preferred frame ($u^0=1$) for $m=1$. The Casimir force is defined by
($F=-dE_c/da$). For $F>0$ the force is repulsive, while for $F<0$ force is
attractive.

Energy as well as the force (Figs~4,~5), shows quasioscillatory behaviour
and changes sign many times. Force is not a differentiable function of
length. There are some `jumps', related to the restriction for the momentum
of particles (Eq.~\ref{eq:pcon}).
The field in a finite length $a$, is composed of modes, which correspond to
the momenta $k_n={\pi}n/a$ (Eq.~\ref{eq:fiop}). These momenta decrease
with the increasing of distance $a$ until they cease to satisfy condition
(\ref{eq:pcon}) and `fall out' from the energy expression.

Period of the observed quasioscillations is equal to $\pi/m$ and for masses
of particles greater than $1eV$ has microscopic magnitude. Hence, for longer
distances $a$ ($\lambda>>1$) quasioscillation might be weakly observable.
So, we examine slow changeable component of the Casimir energy, after
averaging its values in single periods.

\begin{figure}[h]
\begin{center}
\includegraphics[width=7cm]{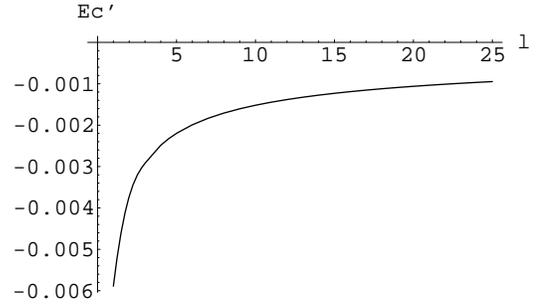}
\caption{Casimir energy after averaging oscilations.}
\end{center}
\end{figure}

The Fig.~6 shows slow changeable component of Casimir energy (as a function
of $\lambda$ parameter). Numerical calculations show, that presented function
behaves like the power function $c\lambda^\alpha$ where $\alpha=-0.5$.

To test the solutions presented on Figs~3-6, we have computed expression in
Eq.~\eqref{eq:CasE} by three other methods, which are not presented
here. There were among others: the method of zeta function and the method of
generalized Abel-Plana formula. All gave the same results.

We realize that the results pressented above seem strange. Nowadays,
tachyons are still hypothetical objects. Maybe it is another reason which
causes that tachyons does not exist. However, this question is still open.
 
\section{Casimir effect for massive field in the preferred frame}

In this section we calculate Casimir energy in the case of the massive field
($m^2>0$) using the absolute synchronization scheme. Our purpose is the
comparison of the results obtained in previous paragraph for tachyons and
the well known ones for massive particles \cite{SzescA, Szosty, Siodmy}.
Analogously as on last section we consider only one dimensional, scalar
field.

The formula on the square of four-momentum for massive particle
\begin{equation}
\label{eq:warC}
k_{\alpha}k^\alpha=+m^2,
\end{equation}
leads to the following dependence of the energy $k_0$ on momentum
$\vec{\underline{k}}$:
\begin{equation}
k_0=(u^0)^{-1}\biggl[-\vec{u\underline{k}}+\sqrt{(
\vec{u\underline{k}})^2+\vec{\underline{k}}^2+m^2}\biggr].
\end{equation}
Lagrangian density of field can be written as
\begin{multline}
\mathcal{L}=\tfrac{1}{2}(g^{\alpha\beta}\partial_\alpha\phi\partial_\beta\phi
-m^2\phi^2)=\\=\tfrac{1}{2}\biggl[(u^0)^2(\partial_0\phi)^2+
u^0u^1(\partial_0\phi\partial_1\phi+\partial_1\phi\partial_0\phi)+\\
-(\partial_1\phi)^2-m^2\phi^2\biggr].
\end{multline}
Scalar, one dimensional and massive field, similarly to the tachyonic
case (Eq.~\ref{eq:fiop}), can be expressed by
\begin{equation}
\phi(x,u)=\frac{1}{\sqrt{2\pi}}\int d\mu'(k)(e^{ikx}a^+(k)+e^{-ikx}a(k)),
\end{equation}
where the invariant measure $d\mu'(k)$ takes now the form
\begin{equation}
\label{eq:mi2}
d\mu'(k)=dk_0dk_1\Theta(k^0)\delta(k^2-m^2)=
dk_0dk_1\frac{\delta(k^0-\omega'_k)}{2\omega'_k},
\end{equation}
and the coefficient $\omega'_k$ is given by
\begin{equation}
\omega'_k=\sqrt{k_1^2+(u^0m)^2}.
\end{equation}

(Index `prime' is  added to differentiate expressions from their tachyonic
equivalents.)

The operator $\phi$, with the measure $d\mu'(k)$ (Eq.~\ref{eq:mi2}),
can be written as
\begin{equation}
\phi(x,u)=\frac{1}{\sqrt{2\pi}}\int_{-\infty}^\infty\frac{dk_1}{2\omega'_k}
(e^{ikx}a^+(k_1,u)+e^{-ikx}a(k_1,u)).
\end{equation}
Contrary to the case of the tachyonic field, we have a full range of
integration because the relation (\ref{eq:warC}) does not lead to any
restrictions for momentum values.

In this section we compare the Casimir energy for massive field with results
from Section III and again we consider only Casimir effect in prefered frame
($u^0=1$).

Hamiltonian of the field in a finite region of length $a$ with boundary
condition (\ref{eq:bkon}) can be expressed by
\begin{equation}
\label{eq:hamc}
H=\frac{1}{2}\int_0^adx\bigl[(\partial_0\phi)^2+(\partial_1\phi)^2+
m^2\phi^2\bigr].
\end{equation}
Fourier decomposition of $\phi$ is defined by
\begin{align}
\label{eq:phip}
\phi(x^0,x^1)=\frac{i}{\sqrt{a}}\sum_{n=-\infty}^\infty
\frac{\sin{k_nx^1}}{2\omega'_k}[e^{-ik_0x^0}a_n^+-e^{ik_0x^0}a_n],
\end{align}
where $k_n={\pi}n/a$. Operators $a_n$ and $a_n^+$ fulfill the following
commutation relations:
\begin{align}
[a_n,a_{n'}]&=[a_n^+,a_{n'}^+]=0,\\
[a_n,a_{n'}^+]&=2\delta_{n,n'}{\omega'_n}.
\end{align}
On inserting $\phi$ from Eq.~\eqref{eq:phip} and its derivatives into
Eq.~\eqref{eq:hamc} we get (after integrating with the respect to $x^1$)
the Hamiltonian in the form:
\begin{multline}
H=\sum_{n=-\infty}^\infty\tfrac{1}{16}{\omega'}_k^{-2}
(k_0^2+m^2+k_n^2)(a_n^+a_n+a_na_n^+)=\\
=\sum_{n=-\infty}^\infty\tfrac{1}{8}(a_n^+a_n+a_na_n^+),
\end{multline}

Now, the vacuum energy is given by
\begin{equation}
E=\langle{0}{\vert}H{\vert}0\rangle=
\tfrac{1}{2}\sum_{n=-\infty}^\infty\sqrt{k_n^2+m^2},
\end{equation}
After the change of the summation limit, we finally obtain
\begin{equation}
E(a)=\sum_{n=1}^{\infty}\sqrt{\biggl(\frac{\pi{n}}{a}\biggr)^2+m^2}.
\end{equation}
Renormalized Casimir energy (a difference between a vacuum energy in two
configurations from Fig.1) takes the form
\begin{equation}
E_c(a)=\frac{\pi}{a}\biggl(\sum_{n=0}^\infty\sqrt{n^2+\lambda^2}-
\int_0^\infty\sqrt{n^2+\lambda^2} \biggr),
\end{equation}
where $\lambda=am/\pi$.

Using the Abel-Plana formula (\ref{eq:ABF}) we finaly receive
\begin{equation}
\label{eq:ecklas}
E_c(a)=-\frac{\pi}{a}\int_\lambda^\infty\frac{\sqrt{n^2-\lambda^2}}
{\exp(2\pi{n})-1}dn.
\end{equation}
The expression in Eq.~\eqref{eq:ecklas} is the well known formula,
describing the Casimir effect for massive field ($m^2>0$), in standard
procedure of synchronization (See \cite{Szosty, Siodmy}).

Below we present Figs~7-9, which are counterparts of the Figs~3-5. The
Casimir energy as a function of $m$, the Casimir energy as a function of
$a$ and Casimir force are shown respectively. 

\begin{figure}[h]
\begin{center}
\includegraphics[width=7cm]{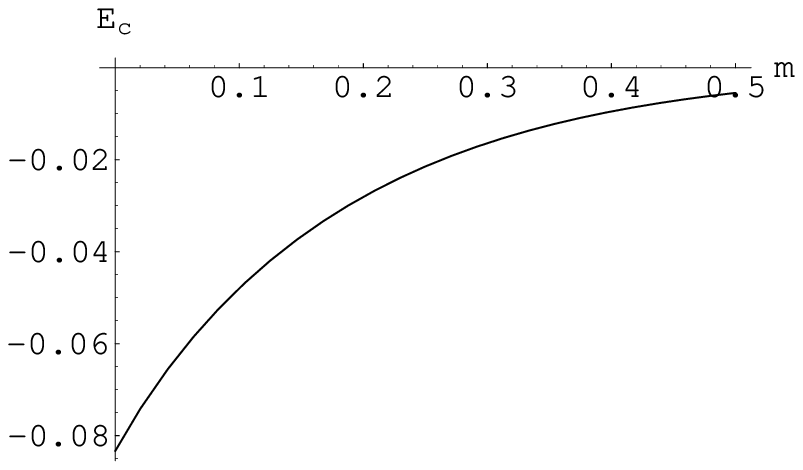}
\caption{Casimir energy for massive field as a function of mass $m$,
corresponding to $a=\pi$.}
\includegraphics[width=7cm]{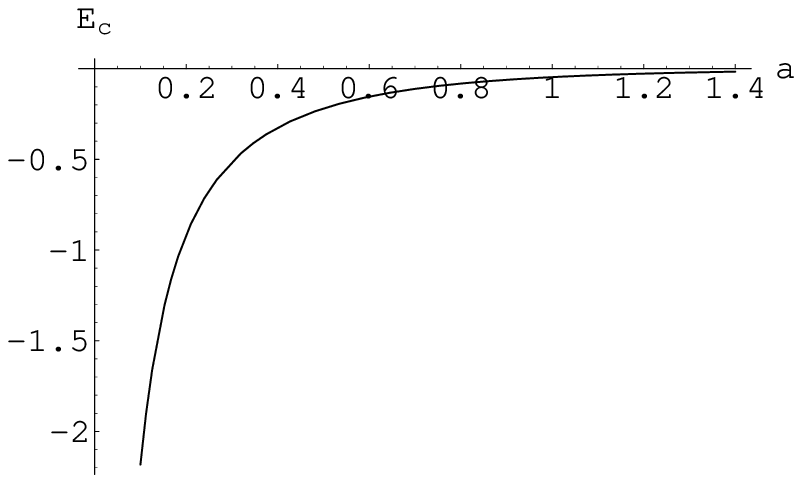}
\caption{Casimir energy for massive field as a function of the distance $a$,
 corresponding to $m=1$.}
\end{center}
\end{figure}

\begin{figure}[h]
\begin{center}
\includegraphics[width=7cm]{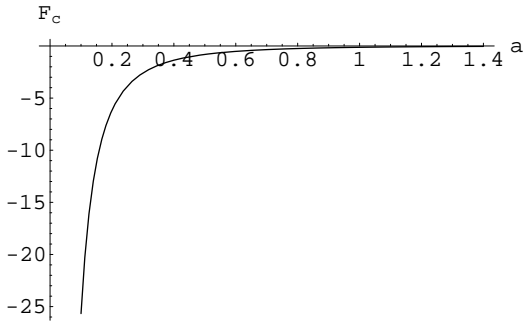}
\caption{Casimir force for massive field as a function of the distance $a$,
 corresponding to $m=1$.}
\end{center}
\end{figure}

\section{Casimir effect for tachyons in nontrivial topology}

In ths section, we again consider the one dimensional, scalar, tachyonic
field in the limited region of length $a$ (Fig.~1). But now we impose
boundary conditions
\begin{equation}
\label{eq:bkon2}
\phi(x^0,0)=\phi(x^0,a),\text{  }\partial_1(x^0,0)=\partial_1(x^0,a),
\end{equation}
which describe the identyfication of the boundary points $x^1=0$ and
$x^1=a$. In this section we consider Casimir energy in arbitrary frame
which is moving relative to preferred frame with fourvelocity $u$.

Taking into account the boundary condition Eq.~\eqref{eq:bkon2} we obtain
the Fourier expression of the operator $\phi$ such that
\begin{align}
\label{eq:fiop2}
\phi(x^0,x^1)=\frac{1}{\sqrt{a}}\sum_{k_n^2>(mu^0)^2}
\frac{1}{2\omega_k}\bigl[e^{ikx}a_n^++e^{-ikx}a_n\bigr],
\end{align}
where $k_n={2\pi}n/a$. In this case Hamiltonian of tachyonic field can be
expressed as
\begin{equation}
\label{eq:ham2}
H=\frac{1}{2}\int_0^adx\bigl[(u^0)^2(\partial_0\phi)^2+(\partial_1\phi)^2-
m^2\phi^2\bigr].
\end{equation}
On inserting operator $\phi$ from Eq.~\eqref{eq:fiop2} with its derivatives
into Eq.~\eqref{eq:ham2} and integrating with respect to variable $x^1$, we
find
\begin{multline}
H=\sum_{k_n^2>(mu^0)^2}\tfrac{1}{8}\omega_k^{-2}
((u^0k_0)^2-m^2+k_n^2)(a_n^+a_n+a_na_n^+).
\end{multline}
If we calculate vacuum energy as an expectation value of $H$, we get
\begin{multline}
\label{eq:vacE2}
E=\langle{0}{\vert}H{\vert}0\rangle=\\=\tfrac{1}{2}(u^0)^{-2}
\sum_{k_n^2>(mu^0)^2}\biggl(-u^0u^1k_n+\sqrt{k_n^2-(u^0m)^2}\biggr),
\end{multline}

After changing the limits of summation, Eq.~\eqref{eq:vacE2} can be written as
\begin{equation}
\label{eq:VacEn2}
E(a,u)=(u^0)^{-2}\sum_{n=n_0}^{\infty}\sqrt{\biggl(\frac{2\pi{n}}{a}\biggr)^2
-(mu^0)^2},
\end{equation}
where $n_0=[mu^0a/2\pi]$. (Function $[x]$ gives the smallest integer
greater than or equal to $x$.) 

The renormalized Casimir energy (a difference between vacuum energy in the
two configurations shown in Fig.~2.) is given by
\begin{multline}
E_c(a,u)=(u^0)^{-2}\\
\times\biggl(\sum_{n=n_0}^{\infty}\sqrt{\bigl(2\pi{n}/a\bigr)^2-
(mu^0)^2}-\frac{a}{2\pi}\int_{mu^0}^\infty{dk}\sqrt{k^2-(mu^0)^2}\biggr).
\end{multline}
Introducing parameter $\lambda(u)=amu^0/2\pi$ and introducing the new
variable of integration $n=ak/2\pi$ we get
\begin{equation}
\label{eq:CasE2}
E_c(a,u)=\frac{2\pi}{a(u^0)^2}\biggl(\sum_{n=n_0}^\infty\sqrt{n^2-
\lambda^2(u)}-\int_\lambda^\infty{dn}\sqrt{n^2-\lambda^2(u)}\biggr),
\end{equation}
where $n_0=[amu^0/2\pi]=[\lambda]$. The expression (\ref{eq:CasE2}) is equal
to expression (\ref{eq:CasE}) multiplied by factor $2(u^0)^{-2}$. Hence,
in the preferred frame ($u^0=1$), energy (\ref{eq:CasE2}) is two times
larger than energy (\ref{eq:CasE}), showed in Figs~3,4.
Since the dependence on $u^0$ enters to the expression \eqref{eq:CasE2} only
through parameter $\lambda$ and factor $\pi/a(u^0)^2$, therefore for other
values of $u^0$ (in moving frame relative to preferred frame) the
Casimir energy function will have the same character, only its horizontal
and vertical scale will change.

\section{Comparison of results for tachyonic and other fields-conclusions} 

Let us compare the results for tachyonic field (Section III) with the
results obtained for the massive ones (Section IV). In the case of the
massive field $m^2>0$ we get monotonic and differentiable dependence of
energy and force (as a functions of distance). Force is always attractive
and smoothly goes to zero when the distance approaches infinity.

In the tachyonic case the situation is completely different. Indeed, there is
no monotonicity at obtained expressions. Energy as well as force,
show quasi-oscillatory behaviour and change own sign many times. Furthermore,
the force is not a differentiable function of distance. There are some
`jumps', where force goes to infinity.

\begin{figure}[h]
\begin{center}
\includegraphics[width=7cm]{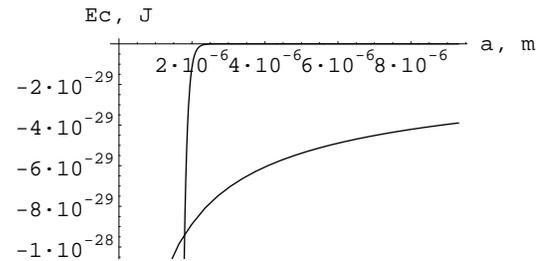}
\caption{Casimir energy for tachyonic and massive fields in preferred frame.}
\end{center}
\end{figure}

Let us compare the behaviour of the Casimir energy as a function of the
length $a$. The Fig.~10 shows Casimir energy for the tachyonic and massive fields.
This figure is made in SI units for mass of particle $m=1eV$.
First (upper) curve shows Casimir energy for massive field. In this case
Casimir energy decay exponentialy with distance $a$. Second (lower) curve
shows Casimir energy for the tachyonic field (averaged over oscillations).
As we mentioned previously, in this case energy behaves like a power function.

For full comparison, we present on the Fig.~11 Casimir energy for massless
field. This figure is made in the same units as Fig.~10. In case of the one
dimensional, massles field, formula for Casimir energy takes well known
form: $E_c(a)=-\pi/12a$ (See \cite{Szosty}).

The behaviour of Casimir energy (averaged over oscillations)
for tachyonic case is more similar to massless field than to massive
field. In both cases (tachyonic and massless) we have slow decay of the Casimir
Energy with distance $a$. Contrary, in case of massive field, Casimir energy
decay fast (exponentialy) with distance $a$.

\begin{figure}[h]
\begin{center}
\includegraphics[width=7cm]{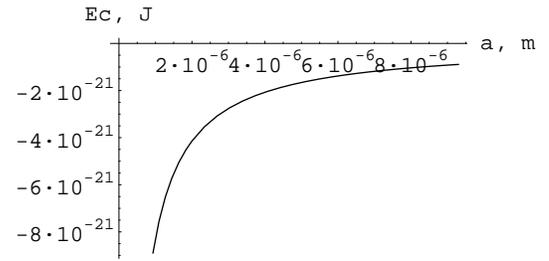}
\caption{Casimir energy for massles field in preferred frame.}
\end{center}
\end{figure}

\end{document}